\documentclass[prl,aps,showpacs,groupedaddress,superscriptaddress,twocolumn,longbibliography]{revtex4-1}

\usepackage[utf8x]{inputenc}
\usepackage{color}
\usepackage{bbm}

\usepackage{nicefrac}
\usepackage{amsfonts,amsmath,amssymb,stmaryrd}
\usepackage{braket}
\usepackage[english]{babel}

\usepackage{tabularx}
\usepackage{multirow} 
\usepackage{array}  
\usepackage{hhline}

\usepackage{graphicx}
\usepackage{subfigure}  
\usepackage{bm}	
\usepackage{bbm}	
\usepackage{hyperref}
\usepackage[pdftex]{epsfig}
\usepackage{mathrsfs}
\usepackage{verbatim}
\usepackage{centernot}

\usepackage{cancel,ifthen}

\usepackage{todonotes}
\presetkeys{todonotes}{size=\tiny}{}

\InputIfFileExists{gitinfo-latex.inc}{}{}
\usepackage[acronym,nomain]{glossaries}

\usepackage{units}
\usepackage{upgreek}

\renewcommand{\vec}[1]{\mathbf{#1}}

\newcommand{\idx}[1]{^{\,}_{#1}}	
\newcommand{\itx}[1]{\idx{\text{#1}}}

\def\d{{\rm d}}
\def\H{\hat{\mathcal{H}}}		
\renewcommand{\a}[1]{\hat{a}\i{#1}}					
\newcommand{\ad}[1]{\hat{a}^{\dagger}_{#1}}		

\renewcommand{\b}[1]{\hat{b}\i{#1}}	

\def\qc1{q_c^{(1)}}
\def\kb{k_{\text{B}}^{\,}}			

\def\v0{\vec{0}}

\def\gbb{g}
\def\gBB{g}
\def\gib{g_\text{IB}}
\def\gIB{\gib^{\,}}
\def\aBB{a_\text{BB}}
\def\aIB{a_\text{IB}}

\def\Tc{T\itx{c}}   
\def\TBEC{T}        
\def\TI{T_{\rm I}}  

\def\mI{m_{\rm I}}  
\def\mB{m_{\rm B}}  
\def\mP{m^*} 	    

\def\p{\hat{\vec{p}}}				
\def\pc{p_{\text{c}}}
\def\k{\vec{k}}						

\def\GammaA{\Gamma^{sp}_{1ph}} 
\def\GammaB{\Gamma^{T}_{1ph}} 

\def\GammaC{\Gamma^{sp}_{2ph}} 
\def\GammaD{\Gamma^{T}_{2ph}} 

\renewcommand{\l}{\left(}
\renewcommand{\r}{\right)}

\renewcommand{\H}{\hat{\mathcal{H}}}

\renewcommand{\a}{\hat{a}}

\renewcommand{\b}{\hat{b}}
\renewcommand{\ad}{\hat{a}^\dagger}

\newcommand{\rh}{\hat{\rho}}

\renewcommand{\vec}[1]{\bm{#1}}

\newcommand{\cmnt}[2][NoInPuT]{\ifthenelse{\equal{#1}{NoInPuT}}{}{{\color{red}\sout{#1}}} {\color{blue} #2}}

\begin{document}
	
	\title{Prethermalization in the cooling dynamics of an impurity in a BEC}
	
	\author{Tobias Lausch}
	\author{Artur Widera}
	\author{Michael Fleischhauer}
	\affiliation{Department of Physics and Research Center OPTIMAS, University of Kaiserslautern, 67663 Kaiserslautern, Germany}

	\pacs{...}
	
	\date{\today}
	
	\begin{abstract}
		We discuss the cooling dynamics of heavy impurity atoms in a Bose-Einstein condensate (BEC) by emission of Cherenkov phonons from scattering with the	condensate.  In a weakly interacting, low-temperature condensate the superfluidity of the condensate results in a separation of time-scales of the thermalization dynamics.  Pre-thermalized states are formed with distinct regions of impurity momenta determined by the mass ratio of impurity and BEC atoms. This can be employed to detect the mass renormalization of the impurity upon the formation of a polaron and paves the way to preparing non-equilibrium impurity-momentum distributions.
	\end{abstract}
	
	\maketitle

	\emph{Introduction.--}
	Cooling atomic quantum gases down to ultralow temperatures has opened a window to experimental exploration of quantum phenomena \cite{Leggett2001,Weiner1999,Pitaevskii2016}. Beyond probing of ground state properties of quantum objects, quantum gases offer the possibility to induce and study non-equilibrium and relaxation dynamics. Tight control over trapping potentials has allowed shedding light onto the non-equilibrium dynamics of integrable systems, which owing to the large number of conserved quantities can stay at a non-thermal steady state  \cite{Kinoshita2006, Hofferberth2007}, described
	by a generalized Gibbs ensemble \cite{Rigol2008}. Weak integrability-breaking perturbations will eventually lead to thermalization, but a separation of time scales can give rise to long-lived prethermalized states \cite{Berges2004,Gring2012}. While much attention is payed to nearly integrable quantum systems, also the thermalization dynamics of non-integrable systems can be non-trivial including long-lived
	prethermalized states resulting from dynamical constraints.
	 Understanding here the microscopic details will help elucidating open questions of quantum thermalization. 
	A powerful tool to trace this microscopic thermalization dynamics is the immersion of impurities in a quantum gas \cite{Hewson1997,Gorshkov2010,Chevy2010,Olf2015}. Particularly, for the paradigmatic system of single impurities in a Bose-Einstein condensate (BEC), the momentum dependent transition to superfluid dynamics is expected to have profound impact on the thermalization. In the present letter we show that, even though the system is non-integrable, this can lead to a separation of time scales between a fast relaxation into a prethermalized state and the eventual approach of thermal equilibrium.
	Moreover, when the impurity is decelerated below the critical momentum of superfluidity, polaronic quasiparticles form \cite{Mathey2004,Rath2013,Li2014a,Levinsen2015,Ardila2015,Ardila2016,Shchadilova2016a}, which has been observed for strong coupling recently \cite{Hu2016,Jorgensen2016}. The polaronic modifications of the quantum state can alter the prethermalized state. This may allow access to properties of these quasiparticles such as the polaronic mass renormalization already for small modifications.

	\begin{figure}[htb]
		\begin{center}
			\includegraphics[width=0.48\textwidth]{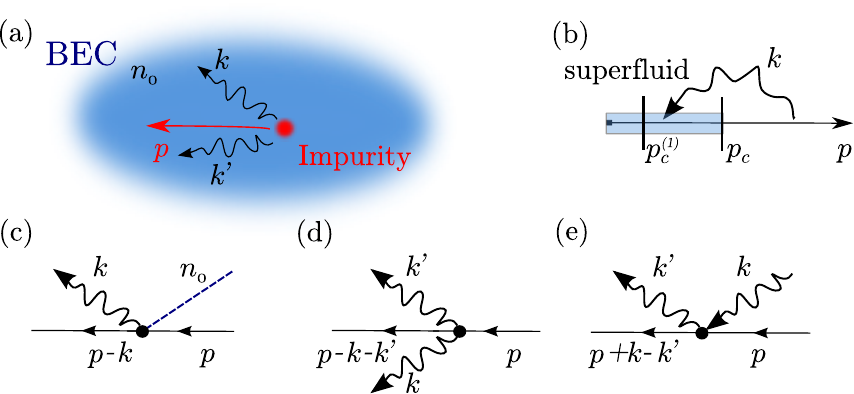}
		\end{center}
		\caption{
			(a) We consider impurity atoms inside an ultracold BEC of density $n_0$,  initially in a thermal state at high temperature.
			The emission of Cherenkov phonons  leads to dissipation of kinetic energy of the impurity and thus to cooling (b).
			The vertices corresponding to spontaneous phonon scattering for one/two scattering partner are shown in (c-e).
			Two phonon scattering which depicted in (d,e) leads to corrections to the standard Fröhlich model.
		}
		\label{fig:Setup}
	\end{figure}

	In this letter we provide a microscopic description of the cooling dynamics of a single mobile impurity immersed in a three-dimensional (3D) superfluid BEC, see Fig.~\ref{fig:Setup} (a). 
	For impurity momenta larger than the Landau critical momentum, deceleration and cooling of the impurity is achieved by Cherenkov-type emission of high energy Bogoliubov excitations, a mechanism previously suggested in Refs.\cite{Daley2004,Griessner2006,Griessner2007} and explored experimentally in Refs.\cite{Scelle2013,Chen2014}. In contrast to these earlier works we theoretically consider a scenario not involving lattice potentials and which is experimentally easily implementable \cite{Hohmann2017}.
	When the impurity mass exceeds the boson mass, the superfluid nature of the bath leads to different critical momentum regions, see Fig.~\ref{fig:Setup} (b). For a weakly 
	interacting BEC, scattering processes into and out of these regions will occur on different time scales giving rise to prethermalization. Scattering processes involving two thermal phonons, see e.g. Fig.~\ref{fig:Setup} (e) eventually lead to a thermalization of the impurity. On shorter time scales, relevant to prepare and observe the prethermalized state, we find that one-phonon processes which can only polulate momentum states above a certain critical value, dominate over two-phonon processes. By contrast, two-phonon terms typically dominate the dynamics of two- and one-dimensional systems on all time scales, as we discuss elsewhere \cite{Lausch2017b}.
		
	Increasing the interaction strength between an impurity with sufficiently low momentum and the BEC leads to polaron dressing. While the polaronic binding energy is readily accessible by radiofrequency spectroscopy \cite{Koschorreck2012,Schmidt2011,Shashi2014RF},
	 other polaronic properties such as quasi-particle weight or effective mass must be deduced by other means. We show that the value of the critical momentum for impurity scattering is sensitive to small modifications of the impurity mass by phonon dressing.

	\emph{Prethermalized impurities in ultracold BECs.--}
	In the following we consider a single impurity atom (mass $\mI$) which is immersed in a BEC of a second, bosonic species of atoms (mass $\mB$). We model their mutual interaction by a local contact potential with strength $\gib$. 
	
	The atoms of the condensate are assumed to weakly interact with each other with interaction strength $\gbb$. In this case the homogeneous BEC can be described using the Bogoliubov approximation, where the original boson operators $\b_{\vec{k}}$ are related to phonons $\a_{\vec{k}}$ by $\b_{\vec{k}} = \a_{\vec{k}} \cosh \theta_k - \ad_{-\vec{k}} \sinh \theta_k$, see e.g. \cite{Pethick2008}, and $\tanh^2\theta_k=\left(\frac{k^2}{2 \mB} + n_{0}\, \gBB-\omega_k\right)
	\left(\frac{k^2}{2 \mB} + n_{0}\, \gBB+\omega_k\right)^{-1}$.  Up to a constant energy offset this leads to the Hamiltonian in $d$ spatial dimensions \cite{Grusdt2015Varenna} 	
	\begin{eqnarray}
		&&\H = \frac{\hat{\vec{p}}^2}{2\mI} + \int \!\! d^d k \left[ \omega_k \ad_{\vec{k}} \a_{\vec{k}} +  \frac{\gib n_0^{1/2}}{(2 \pi)^{d/2}}  W_k e^{i \vec{k} \cdot \hat{\vec{r}}} \l \ad_{\vec{k}} + \a_{\vec{k}}  \r \right]\nonumber\\ 
		&&\enspace +\frac{\gib}{2 (2 \pi)^d}  \int \!\! d^dk \int \!\! d^dk'   \biggl[ \l W_k W_{k'} + W_k^{-1} W_{k'}^{-1} \r \ad_{\vec{k}} \a_{\vec{k}'}\label{eq:HBogoPolaron} \\
		&&+ \frac{1}{2} \l W_k W_{k'} - W_k^{-1} W_{k'}^{-1} \r \l \ad_{\vec{k}} \ad_{-\vec{k}'} + \a_{-\vec{k}} \a_{\vec{k}'} \r \biggr] e^{i \l \vec{k} - \vec{k}' \r \cdot \hat{\vec{r}}}.\nonumber
	\end{eqnarray}
	Here we consider trapped systems in $d=3$ spatial dimensions, but will disregard the effects of the trapping potential to the Bogoliubov excitations. 
	$\hat{\vec{p}}$ ($\hat{\vec{r}}$) are the impurity momentum (position) operators and $n_0$ denotes the density of the condensate. Here and in the following we set $\hbar=1$ and $k_B=1$. Finally we defined $W_k =[k^2 \xi^2 / (2 + k^2 \xi^2)]^{1/4}$ and the Bogoliubov dispersion reads $\omega_{\vec{k}} = c k ( 1 +  k^2 \xi^2  / 2 )^{1/2}$, with $c=\sqrt{\gBB n_{0}/\mB}$ being the speed of sound and $\xi = 1/\sqrt{2 \mB \gBB n_{0}}$ the healing length of the BEC with density $n_0$. 
	The term in the first line of Eq.~\eqref{eq:HBogoPolaron} corresponds to the Fröhlich Hamiltonian \cite{Froehlich1954}, known to describe polarons in solid-state systems. The Ginzburg radiation \cite{Ginzburg1996} resulting from the Fröhlich term has been investigated for particles with internal structure in \cite{Marino2017}.
	The terms in the last two lines include two-phonon scattering terms, which in general need to be taken into account in a BEC \cite{Rath2013,Li2014a,ChristensenPRL2015}.
	
	In the following we want to study the non-equilibrium dynamics of Hamiltonian \eqref{eq:HBogoPolaron}, starting from a thermal state of the bosons and a thermal momentum distribution of the impurity centered around a value $\vec{p}_0$. When their temperatures are given by $\TBEC$ and $\TI$ respectively, the density matrix before BEC-impurity interaction is $\rh(0) = \exp [- (\hat{\vec{p}}-\vec{p}_0)^2/2\mI \TI] \otimes  \exp [- \int d^dk ~ \omega_k \ad_{\vec{k}} \a_{\vec{k}}/ \TBEC]$. We consider the case when the temperature of the BEC is well below the critical temperature for condensation $\Tc=2\pi n_0^{2/3}/(\mB \zeta(3/2)^{2/3})$, i.e. $\TBEC \ll \Tc$ and assume a heavy impurity, i.e. $\mI > \mB$.
	
	\begin{figure}[htb]
		\begin{center}
			\includegraphics[width=0.48\textwidth]{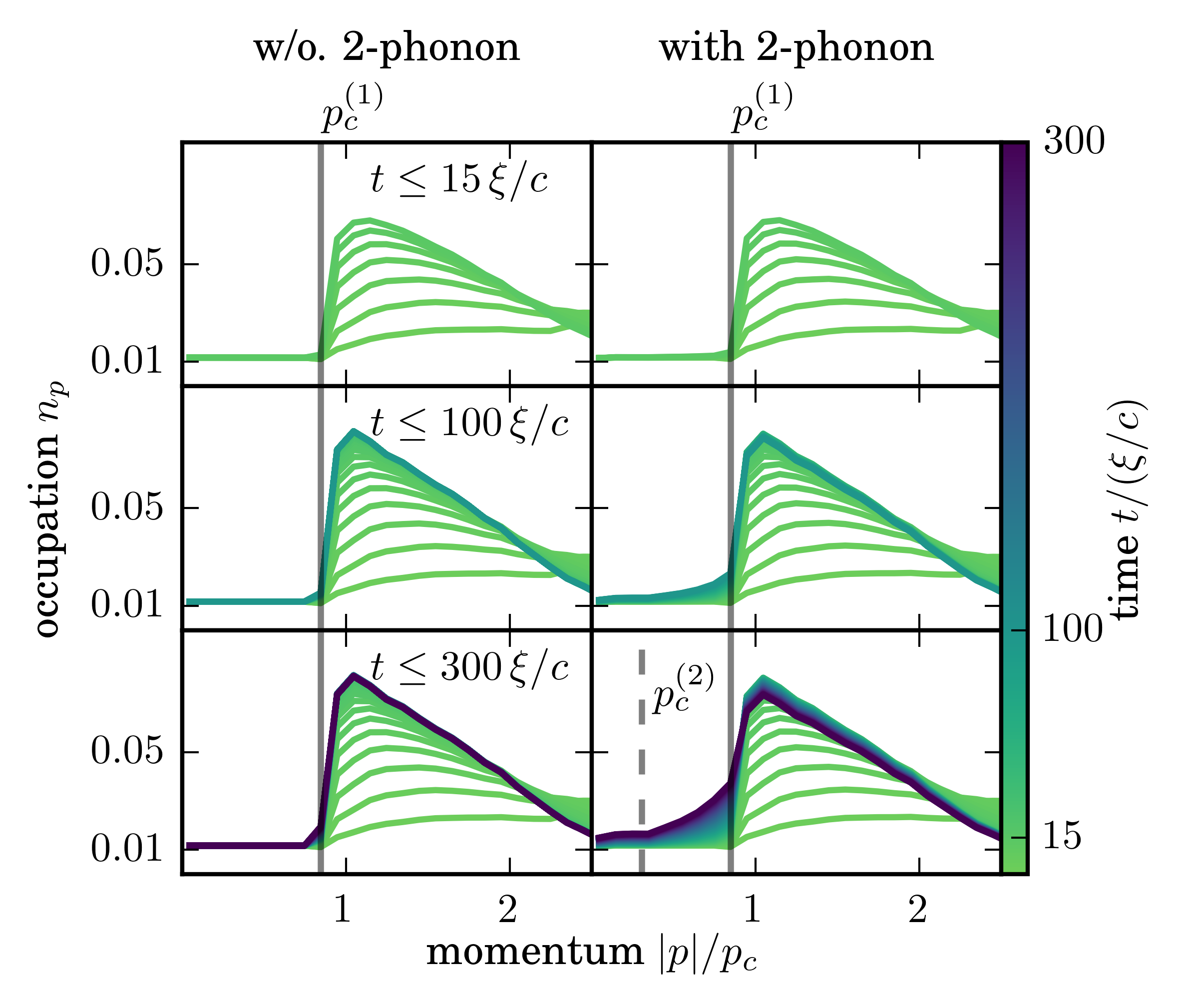}
		\end{center}
		\caption{Numerical simulation of the impurity momentum distribution
	 when starting from a thermal state with $\vec{p}_0=0$ and $\TI= 10 \Tc$ and a low BEC temperature $\TBEC=0.1 \Tc$. Here, we assume a peak density of $n_{0} = 10 /\xi^3$, $\gib = 1$, and a mass ratio $\mI / \mB = 87 / 39$ corresponding to a Rubidium (Rb) impurity in a Potassium (K) BEC. Momenta are given in units of the Landau critical momentum $\pc = \mI c$.
			\textit{Left column:} Dynamics including only single-phonon processes. \textit{Right column:} Dynamics including one- and two-phonon processes. 
			Vertical lines indicate emerging critical momenta $\pc^{(1)} = c\sqrt{\mI^2-\mB^2}$ (solid vertical line) and $\pc^{(2)} = c\sqrt{\mI^2-4 \mB^2}$ (dashed vertical line).
			}
		\label{fig:Dynamics}
	\end{figure}
	
	We use a master equation to describe the dynamics of the impurity-density matrix $\rh_{\rm I}(t)$, which can be derived by integrating out thermal phonons and employing the Born-Markov approximation. The Born approximation neglects higher-order scattering contributions and is valid for weak impurity-condensate interactions $\gIB$. In this way we obtain a linear Boltzmann equation for the momentum distribution $n_{\vec{p}}$ of the impurity, which has different contributions (complete expressions
	are given in the supplementary material.~\cite{Supplementary}, see Eqs.~\eqref{eq:Boltzmann1} -\eqref{eq:Boltzmann4}
	\begin{equation}
		\frac{d n_{\vec{p}}}{dt} = \underbrace{\frac{d n_{\vec{p}}}{dt} \Bigr\vert_{\rm sp,1ph} + \frac{d n_{\vec{p}}}{dt} \Bigr\vert_{\rm T,1ph} + \frac{d n_{\vec{p}}}{dt} \Bigr\vert_{\prec}}_{\rm type ~1} +  \underbrace{\frac{d n_{\vec{p}}}{dt} \Bigr\vert_{\times}}_{2}.
		\label{eq:BoltzmannEquationOverview}
	\end{equation}
	The first two terms describe one-phonon emission, including spontaneous (sp) and thermally activated (T) processes. The third term ($\prec$) describes two-phonon creation or annihilation (thermal and spontaneous). These terms (type 1) are illustrated in the diagrams shown in Fig.\ref{fig:Setup} (c,d).
	The last term (type 2) describes the scattering of a phonon at the impurity and is, in a weakly interacting BEC, where quantum depletion can be
	disregarded, and in lowest-order Born approximation in $\gIB$ only relevant at finite temperatures. It is illustrated in  Fig.\ref{fig:Setup} (e). 
	Due to the superfluidity of the BEC spontaneous phonon creation out of the condensate can only take place for impurity momenta $p$ larger than the Landau critical momentum $\pc = \mI c$ and thus these processes cannot lead to a redistribution below $\pc$.  Higher-order scattering will eventually lead to a thermalization of the impurity. 
			
	We have numerically simulated the time evolution of the momentum distribution of the impurity in a homogeneous BEC at finite $T\ll T_c$ using Eq.~\eqref{eq:BoltzmannEquationOverview}. The results in Fig.~\ref{fig:Dynamics} show the impurity momentum distribution $n_p = \int d^3 \vec{p'}  n_{\vec{p'}} \delta(p-|\vec{p'}|) $ of states having a momentum value of $\vert \vec{p} \vert=p$ with color encoded evolution time. 
	One clearly recognizes that the emission of Cherenkov phonons leads to a fast population of momentum states below $\pc$ but only above a certain critical momentum
	\begin{equation}
		\pc^{(1)} =c\,  \sqrt{\mI^2 - \mB^2},
		\label{eq:pc1}
	\end{equation}
	which can be derived from energy-momentum conservation associated with spontaneous single-phonon creation  \cite{Supplementary}. On a much longer time scale
	two-phonon processes, not accounted for in the Fröhlich model, lead to a population of momentum states below $\pc^{(1)}$. Here energy-momentum conservation
	leads to a second characteristic momentum scale for $\TBEC=0$, provided $\mI > 2 \mB$
	\begin{equation}
		\pc^{(2)} =c\, \sqrt{\mI^2 - 4  \mB^2}.
		\label{eq:pc2}
	\end{equation}
	One recognizes from Fig.~\ref{fig:Dynamics} a clear separation of time scales for scattering events populating momenta above $\pc^{(1)}$ versus those populating lower momenta. For $\pc^{(2)}$ such a separation is much less visible however.

	From the Boltzmann equations one can determine the characteristic rates of spontaneous single- ($\GammaA$) and two-phonon ($\GammaC$)
	processes for an initial impurity momentum $p$ 

	\begin{eqnarray}
		\GammaA(p) &=& \gib^2 n_{0}\int\frac{\d^3k}{(2\pi)^{2}}\, W_k^2 \,\delta\left(\epsilon_{\vec{p}} -\epsilon_{\vec{p}-\vec{k}}-\omega_{\vec{k}}\right),
		\label{eq:scattering_1ph_sp}\\
		\GammaC(p) &=& \frac{\gib^2}{2(2\pi)^5} \int\!\d^3k\!\!\int \!\d^3k^\prime \left(\frac{W_{k} W_{k^\prime} -
			W_{k}^{-1} W_{k^\prime}^{-1}}{2}\right)^2\nonumber\\
		&& \times\,\delta\left(\epsilon_\mathbf{p} -\epsilon_{\vec{p}-\vec{k}-\vec{k^\prime}}-\omega_{\vec{k}}-\omega_{\vec{k^\prime}}\right).
		\label{eq:scattering_2ph_sp}
	\end{eqnarray}
	%
	%
	\begin{figure}[t]
		\begin{center}
			\includegraphics[width=0.48\textwidth]{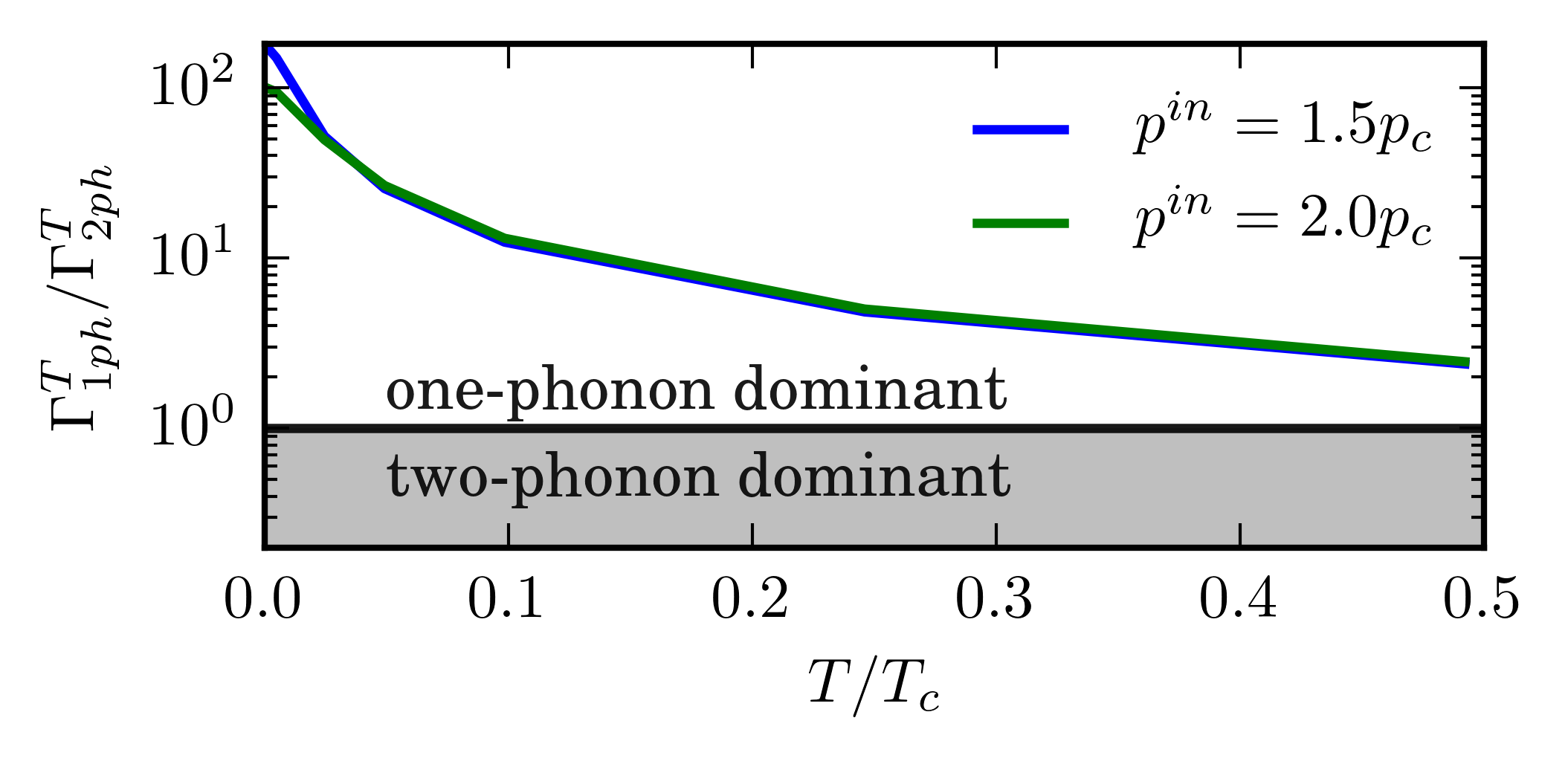}
		\end{center}
		\caption{Ratio of single-phonon to two-phonon scattering rates at finite $T$ for different incoming impurity momenta. We find that scattering is dominated by single-phonon processes in either case, spontaneous ($\TBEC=0$) or thermal. With an increasing BEC temperature the two-phonon scattering rate increases faster than the single-phonon scattering rate, hence two-phonon processes become more important at large temperatures. Here $n_0 = 5$ and $\xi = 1$.
		}
		\label{fig:scattering_rate_ratios}
	\end{figure}
	The relative scaling of the two rates can be estimated as
	\begin{equation}
		\frac{\GammaC}{\GammaA} \sim\,  \frac{1}{2\pi \xi^3 n_0 }.
	\end{equation}
	We have verified this scaling in our numerical simulations. 
	For weak interactions $n_0\xi^3 \gg 1$ and thus spontaneous two-phonon scattering occurs at a much lower rate than single-phonon scattering, which leads to the 
	separation of time scales observed in the numerical simulations.
	
	\emph{Finite Temperature effects.--} Next we discuss the influence of a non-vanishing BEC temperature. While spontaneous scattering results in a fast increase of population in the momentum region $\pc^{(1)}\leq p \leq \pc^{\,}$, thermal scattering can also cause a depopulation of that region. Thermally induced single-phonon scattering can however not lead to impurity momenta below $\pc^{(1)}$. Thus prethermalization prevails also at finite $T$ as long as single-phonon scattering dominates.
	
	Since the two-phonon rate is proportional to the square of the thermal phonon number $\bar{n}_{k}^2$, it increases faster with temperature than the single-phonon rate. At a certain characteristic temperature $T_{\text{2ph}}^{\,}$ there is a crossover between the single- and two-phonon dominated regimes.
	This crossover can be characterized by considering the thermal phonon number at some characteristc phonon momentum $k_{0}^{\,} = 1/\xi \sim \mathcal{O}(\pc^{\,})$.
	The thermal phonon number at $\k_{0}^{\,}$ exceeds unity for  $T > T_{\text{c}^{\,}}) (n_{\text{0}}^{\,}\xi^{3})^{-2/3}$.
	Since for weak interactions $n_{\text{0}}^{\,}\xi^{3} \gg 1$ temperatures sufficiently below the critical temperature are required in order to be determined entirely by spontaneous processes.
	The crossover temperature $T_{\text{2ph}}^{\,}$ can be estimated setting the ratio of finite-$T$ one- and two-phonon scattering rates $\GammaD / \GammaB \sim \bar{n}_{k_{\text{0}}}^{\,}\GammaC / \GammaA = 1$.
	This yields
	\begin{equation}
		\frac{T_{\text{2ph}}^{\,}}{\Tc}\biggr|_{\text{3D}} \approx \frac{\left[\zeta(3/2)\right]^{2/3}}{\sqrt{2}}\left(n_{\text{0}}^{\,}\xi^3\right)^{1/3}
	\end{equation}
	which, for weak interactions, is larger than unity and thus the crossover occurs only above  $\Tc$. This is illustrated in Fig.~\ref{fig:scattering_rate_ratios}, where we have plotted the ratio of rates of single- and two-phonon creation at finite temperature, obtained from numerical integration.
	\emph{Effects of polaronic mass renomalization.--}
	While there is no spontaneous emission of Bogoliubov phonons for impurity momenta $p<\pc$ (sub-sonic regime), the interaction with the condensate phonons leads to the formation of a polaron \cite{Landau1948,Froehlich1954, Cucchietti2006, Tempere2009}. 
	A polaron is a quasiparticle that describes the impurity dressed by a cloud of phonons carried along with the impurity.
	The strength of the
	impurity-BEC coupling is characterized by a dimensionless coupling 	constant $\alpha$ which describes the ratio of
	interaction energy to a characteristic energy scale of phonons \cite{Tempere2009}
	\begin{equation}
		\alpha = \frac{\aIB^2}{\aBB\, \xi},
	\end{equation}
	where $\aIB$ and $\aBB$ are the boson-boson and impurity-boson scattering length, $\gBB = 2\pi \aBB/\mB$ and $\gIB = 2 \pi \aIB/m_{\rm red}$, with $m_{\rm red}=\mI
	\mB/(\mI +\mB)$ being the reduced mass.

\begin{figure}[t]
		\begin{center}
			\includegraphics[width=0.48\textwidth]{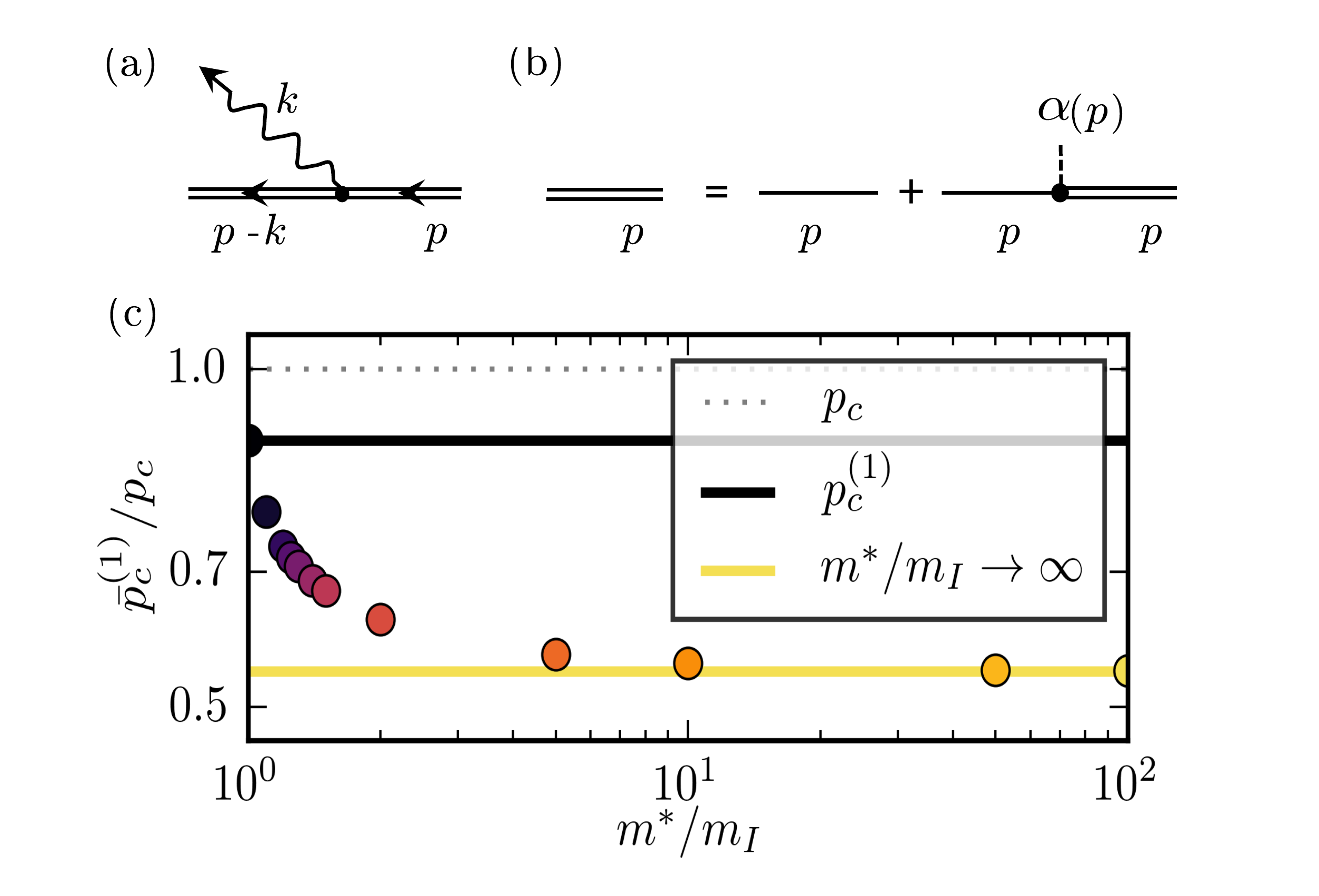}
		\end{center}
		\caption{
			a) Effects of polaronic mass renormalization on the cooling dynamics are taken into account by replacing the bare impurity propagator with the dressed one.
			b) Mean-field approximation to the dressed impurity propagator.
			c) Effect of polaronic mass renormalization to cut-off momentum 
			$\pc^{(1)}$ of single-phonon scattering for mass ratio $\mI/\mB = 87/39$ corresponding to a Rb impurity in a K BEC. The dressing $\mP / \mI$ is color encoded.
			Black solid line depicts no dressing ($\mP / \mI = 1$ ). The yellow solid line is a theoretical extrapolation showing that a finite critical momentum $\pc^{(1)}$ exists even for heavy polarons ($\mP/\mI \to \infty$).
		}
		\label{fig:mass-renorm}
	\end{figure}

	In order to take into account the effects of a polaronic dressing of the impurity to the cooling dynamics, we replace the bare impurity propagator in the scattering diagrams
	of Fig.~\ref{fig:Setup} by the full propagator including self-energy contributions. The corresponding diagram for the single-phonon scattering, which is dominant for short times and
	$n_{0}\xi^3 \gg 1$ is shown in Fig.~\ref{fig:mass-renorm}(a). We evaluate the self energy in mean-field approximation, which is
	justified in the weak-coupling regime.  To this end we transform Hamiltonian \eqref{eq:HBogoPolaron} into 
	the moving frame of the polaron and apply a mean-field shift to the Bogoliubov phonons $U=\exp\left\{\int \d^3 k\, (\alpha_k(p) \hat a_k^\dagger - h.a.)\right\}$.
	The resulting mean-field Hamiltonian is then minimized varying $\alpha_k(p)$. One finds that no solution for $\alpha_k(p)$ exists for $p > p_c\approx \mI c$.
	We thus replace the full impurity propagator by its free propagator for $p > \mI c$ and by a dressed propagator for $p < \mI c$, which could be obtained from 
	solving the mean-field Dyson equation illustrated in Fig.~\ref{fig:mass-renorm}(b).
	The most prominent effect resulting from the formation of the polaron is the renormalization
	of its mass to a value $\mP > \mI$ which is due to the phonon cloud attracted by the impurity \cite{Landau1948}.
	We take this into account by replacing $\mI$ by $\mP$ whenever $p < \mI c$, for details see \cite{Supplementary}.
	The mass renormalization can become rather substantial, when the interaction becomes large, i.e. $\alpha \gg 1$,
	and when the impurity is light, i.e. $\mI \ll \mB$ \cite{Grusdt2016}. 
	Lowest-order perturbation theory in the coupling strength $\alpha$ yields $m^*-\mI \sim \alpha$.
	
	Most interestingly we find that the polaronic mass renormalization affects the cut-off momentum of the phonon scattering terms, Eqs.\eqref{eq:pc1} and \eqref{eq:pc2}.
	Fig.~\ref{fig:mass-renorm}(c) shows the modified critical momentum $\bar{\pc}^{(1)}$ for one-phonon scattering as a function of polaron mass (for details see supplementary material~\cite{Supplementary}). One notices a sizable shift already for moderate mass renormalization.
	In the limit $\mP/\mI \to \infty$ the cut-off approaches the value $\pc^{(1)} = \pc - \mB c\sqrt{\sqrt{4 +(\mI/\mB)^2} - 2}$.  	
	An intriguing perspective based on our findings is to determine the polaronic mass renormalization via measurement of the momentum distribution of the impurity.
	%
	%
	\emph{Summary and Outlook.--}
	Concluding, we have numerically studied the cooling dynamics of individual impurities in a BEC at non-zero temperature, including scattering processes up to second order. We find that the range of accessible final impurity momenta is restricted to values above a critical momentum $p_c^{(1)}$ for short times, while equilibration to a thermal state occurs on a longer time-scale, dominated by two-phonon scattering processes. This time-scale separation leads to a prethermalized quantum state of the impurity as long as energy-redistribution by impurity-impurity interaction can be neglected.
	
	Moreover, the critical momentum obtained depends on the mass renormalization of the polaronic quasi-particle state forming. This is a novel route to experimentally measuring the polaronic mass shift in cold atom experiments.
	
	Importantly, our work paves the way for dynamical quantum state engineering of non-thermal impurity states, for example by shaping its prethermalized momentum distribution. This can be done by operations selective on the impurity's momentum state being fast (slow) compared to the inverse two- (one-) phonon rate.

	\emph{Acknowledgements.--}
	We thank F. Grusdt, K. \"Unl\"ut\"urk and H. Haug for valuable input at different stages of the work. 
	Furthermore we thank Y. Shchadilova, D. Mayer, F. Kindermann, M. Hohmann and F. Schmidt  for stimulating discussions. 
	T.L.~acknowledges financial support from the Carl-Zeiss Stiftung. The work was supported by the Deutsche Forschungsgemeinschaft within the Sonderforschungsbereich (SFB) TR49.

	\section{Supplementary Material}
	\subsection{Boltzmann equation}
	Using the impurity dispersion $\epsilon_{\vec{p}} = \vec{p}^2/2 \mI$ the terms in the Boltzmann equation \eqref{eq:BoltzmannEquationOverview} read:
	\begin{widetext}
		\begin{flalign}
			\frac{d n_{\vec{p}}}{dt} \Bigr\vert_{\rm sp,1ph} &= - \frac{\gib^2 n_0}{(2\pi)^{d-1}} \int d^d{k} ~ W_k^2 \Big[ n_{\vec{p}}  \delta \l \epsilon_{\vec{p}} - \epsilon_{\vec{p} - \vec{k}} - \omega_k \r - n_{\vec{p}+\vec{k}}  \delta \l \epsilon_{\vec{p} + \vec{k}} - \epsilon_{\vec{p}} - \omega_k \r \Big]. \label{eq:Boltzmann1} \\
			\frac{d n_{\vec{p}}}{dt}  \Bigr\vert_{\rm T,1ph} &= - \frac{\gib^2 n_0}{(2\pi)^{d-1}}  \int d^d{k} ~ \bar n_{\vec{k}}(T) W_k^2  \l n_{\vec{p}} - n_{\vec{p}-\vec{k}} \r \Big[   \delta \l \epsilon_{\vec{p}} - \epsilon_{\vec{p} - \vec{k}} - \omega_k \r + \delta \l \epsilon_{\vec{p} - \vec{k}} - \epsilon_{\vec{p}} - \omega_k \r \Big].  \label{eq:Boltzmann2}\\
			\frac{d n_{\vec{p}}}{dt}  \Bigr\vert_{\prec} &=  - \frac{\gib^2}{8 (2\pi)^{2d-1}} \int d^d{k}_1 d^d{k}_2 ~ \l W_{k_1} W_{k_2} - W_{k_1}^{-1} W_{k_2}^{-1} \r^2 \Bigl\{ \l 1 + \bar n_{\vec{k}_1}(T) \r \l 1 + \bar n_{\vec{k}_2}(T) \r \nonumber \\
			& \times \Big[ n_{\vec{p}} \delta \l \epsilon_{\vec{p}} - \epsilon_{\vec{p} -  \vec{k}_1 - \vec{k}_2} -  \omega_{k_1} - \omega_{k_2} \r  - n_{\vec{p}+\vec{k}_2+\vec{k}_1} \delta \l \epsilon_{\vec{p} +  \vec{k}_2 + \vec{k}_1} - \epsilon_{\vec{p}} - \omega_{k_2} - \omega_{k_1} \r   \Big] + \bar n_{\vec{k}_1}(T)  \nonumber \\
			& \times \bar n_{\vec{k}_2}(T) \Big[ n_{\vec{p}} \delta \l \epsilon_{\vec{p}} + \omega_{k_1} + \omega_{k_2} - \epsilon_{\vec{p} + \vec{k}_1 + \vec{k}_2}  \r  - n_{\vec{p}-\vec{k}_2-\vec{k}_1} \delta \l  \epsilon_{\vec{p}} - \omega_{k_2} - \omega_{k_1} - \epsilon_{\vec{p} -  \vec{k}_2 - \vec{k}_1} \r   \Big] \Bigr\}. \label{eq:Boltzmann3}\\
			\frac{d n_{\vec{p}}}{dt}  \Bigr\vert_{\times} &= - \frac{\gib^2}{4 (2\pi)^{2d-1}} \int d^d{k}_1 d^d{k}_2 ~ \l W_{k_1} W_{k_2} + W_{k_1}^{-1} W_{k_2}^{-1} \r^2 \bar n_{\vec{k}_1}(T) \l \bar n_{\vec{k}_2}(T) + 1 \r \nonumber \\
			& \qquad \times \Big[ n_{\vec{p}} \delta \l \epsilon_{\vec{p}} + \omega_{k_1} - \epsilon_{\vec{p} +  \vec{k}_1 - \vec{k}_2} - \omega_{k_2} \r  - n_{\vec{p}+\vec{k}_2-\vec{k}_1} \delta \l \epsilon_{\vec{p}} + \omega_{k_2} - \epsilon_{\vec{p} +  \vec{k}_2 - \vec{k}_1} - \omega_{k_1} \r   \Big] \label{eq:Boltzmann4}
		\end{flalign}
	\end{widetext}
	Here the thermal Bose-Einstein distribution function is given by
	\begin{equation}
		\bar n_{\vec{k}}(T) = \left[\exp \l  \omega_{\vec{k}} / \kb T \r - 1\right]^{-1}.
	\end{equation}

	\subsection{Numerical Modeling}
	
\begin{figure}[htb]
		\begin{center}
			\includegraphics[width=0.48\textwidth]{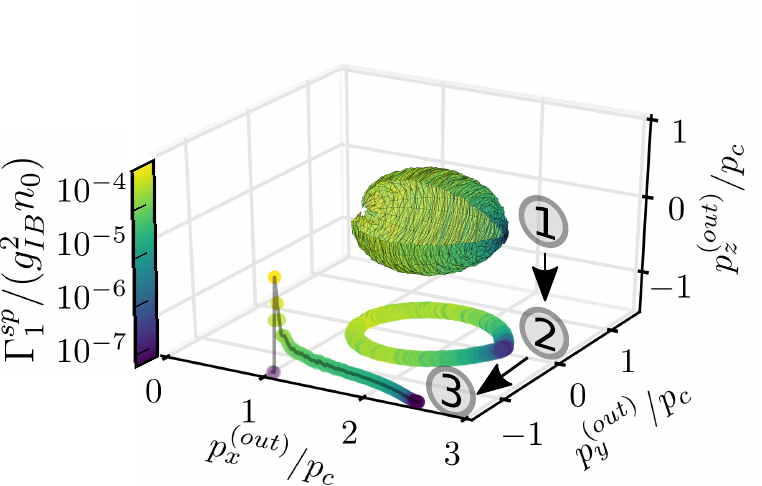}
		\end{center}
		\caption{
			Three-dimensional map of the spontaneous one-phonon scattering rate of different outgoing impurity momenta for an incoming impurity momentum of $\p^{(\textrm{in})}= 2.5 \pc$. The numbers indicate the steps for dimensional reduction to 1D, owing to the symmetry of  scattering.
			}
	\label{fig:numerics_3d}
	\end{figure}

	The simulations are performed by, first, evaluating the delta distribution in Eqs.~\eqref{eq:scattering_1ph_sp}, \eqref{eq:scattering_2ph_sp} for a given incoming impurity moment $p^{(\textrm{in})}$. Considering only one-phonon processes, this results in a large manifold of possible outgoing momenta $\lbrace p^{(\textrm{out})} \rbrace$, forming a sharply defined 3D surface in momentum space, see Fig.~\ref{fig:numerics_3d} for illustration.  Due to radial symmetry, these can be reduced in subsequent steps to eventually 1D by radial integration, allowing for an efficient numerical calculation of scattering rates at moderate numerical effort. We thus obtain an evolution of the impurity state in momentum space, while in our approach the BEC remains a un-perturbed quantum bath.

	For two-phonon processes, the distribution of outgoing momenta is no longer restricted to the 3D surface in momentum space but becomes smeared out to a volume which is however sparsely filled. 
	Clearly, in the limit of weak interactions, $\xi^3 n_0 \gg 1$,
	the scattering rates for momenta in the inner (volume) part 
	of the 3D momentum distribution are much smaller than the scattering rates corresponding to momenta on the surface; thus single-phonon scattering processes in the BEC lead to much faster dynamics than two-phonon scattering processes.
	
	\subsection{Critical Momenta}

	All scattering processes in Eqs.~\eqref{eq:Boltzmann1}-\eqref{eq:Boltzmann4} conserve both, energy and momentum. When an incoming impurity momentum $p^{(\textrm{in})}$ interacts with the bath and creates an excitation, the momentum $k$ is transfered. The resulting momentum $p^{(\textrm{out})}$ then fulfills 
	\begin{align}
		p^{(\textrm{in})} - p^{(\textrm{out})} + k &= 0
		\label{eq:A1_momentum_conservation} \\
		\epsilon_{p^{(\textrm{in})}} - \epsilon_{p^{(\textrm{out})}} - \omega_{k} &= 0\,.
		\label{eq:A1_energy_conservation}
	\end{align}

	Obviously this limits the resulting impurity momenta in 1D to

	\begin{equation}
		p^{(\textrm{out})}(k) = \frac{\mI}{k}\left(\omega_k^{\,} - \epsilon_k^{\,}\right)\,.
		\label{eq:A1_final_momentum}
	\end{equation}

	The resulting momentum $p^{(\textrm{out})}$ has a global minimum when the created excitation and incoming impurity are anti-parallel. Thus we obtain a global minimum for all dimensions $p^{(\textrm{out})}_{\textrm{min}}(k_c^{(1)}) = \pc^{(1)}$ at $k_{c}^{(1)} = \frac{2c^2\mI^2}{p_c^{(1)}}$. This critical momentum $\pc^{(1)} < \pc$, see Eq.~\eqref{eq:pc1} exists for heavy impurities $\mI > \mB$ and is below the Landau critical momentum $\pc$.

	Given the case that the impurity is heavier $\mI > 2 \mB$ the boson mass one finds a second critical momentum $\pc^{(2)}$ for the spontaneous two-phonon scattering $\GammaC$.

\subsection{Polaronic Mass Renormalization}

	The polaronic dressing of the impurity leads to changes of the impurity's interaction with the quantum gas. First, modifications of the energy spectrum due to the polaronic binding energy have been measured in the vicinity of a Feshbach resonance \cite{Hu2016,Jorgensen2016}. Second, the effective mass of the impurity is expected to change due to polaron formation. For our system we thus expect a modification of the critical momentum that an impurity may reach via the resonant scattering processes depicted in Fig. \ref{fig:Setup}.
	We introduce the polaron mass $\mP$ and effectively replace the free impurity propagator in the Eqs.~\eqref{eq:Boltzmann1}-\eqref{eq:Boltzmann4} by a free polaron particle if the impurity's velocity becomes smaller than the speed of sound $c$.
	\begin{align}
		\epsilon_p = 
		\begin{cases}
			p^2 / 2 \mI, & \text{if}\ p>\pc \\
			p^2 / 2 \mP, & \text{otherwise}
		\end{cases}
		\label{eq:polaronic_mass}
	\end{align}
	
\begin{figure}[htb]
		\begin{center}
			\includegraphics[width=0.48\textwidth]{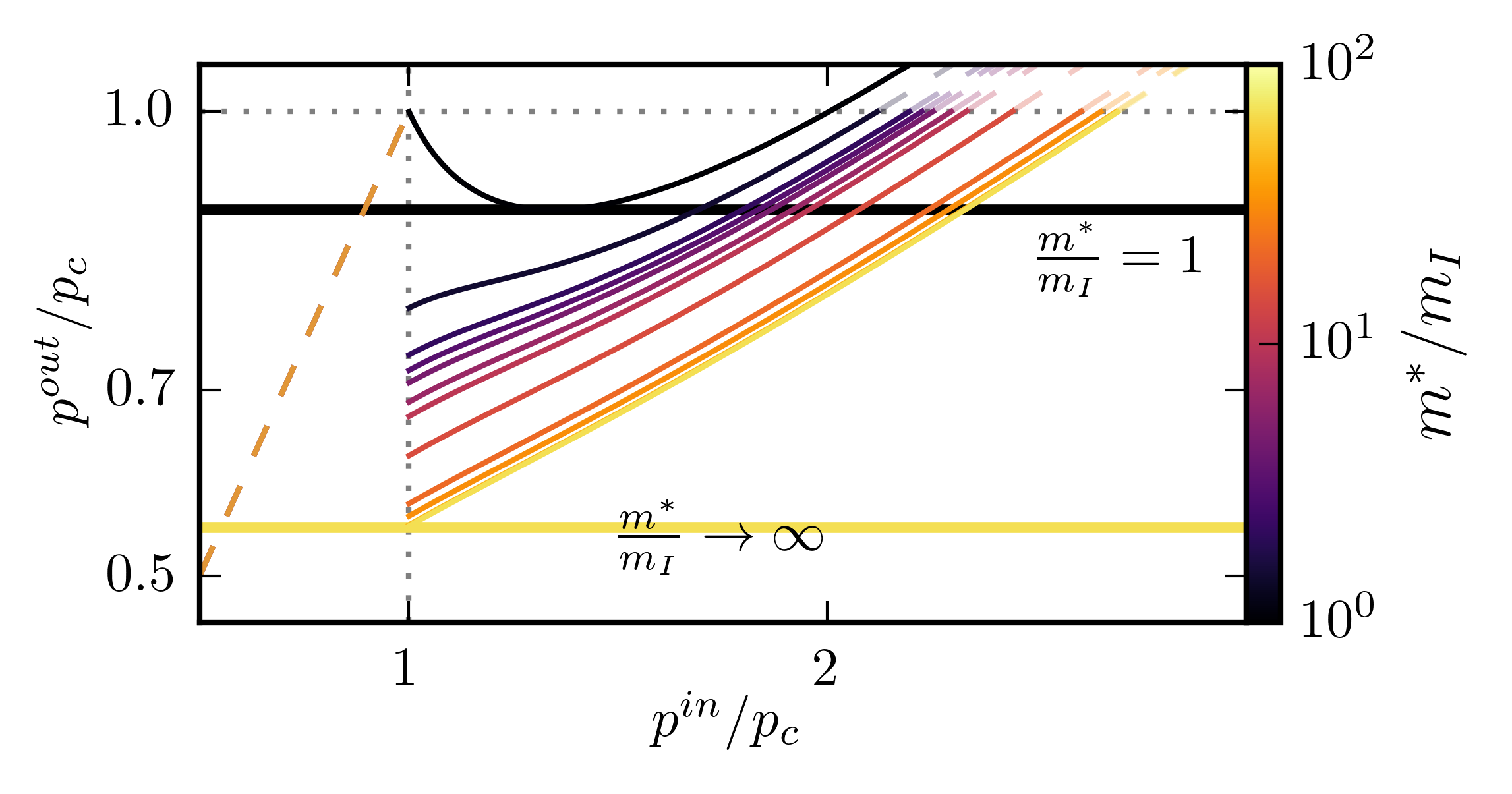}
		\end{center}
		\caption{Possible outgoing impurity momentum $p^{(\mathrm{out})}$ versus incoming impurity momentum $p^{(\mathrm{in})}$ for different effective impurity masses $\mP/\mI$. The bare impurity case (no dressing) is indicated by the black line, while an infitely heavy polaron is indicated by the light yellow line.
			We find no solution of Eq.~\eqref{eq:A1_final_momentum} if the impurity is already dressed by phonons $p^{(\textrm{in})} / \pc \leq 1$.
			We define the minimum of $p^{(\textrm{out})}$ for every dreesing curve as the first critical momentum $\pc^{(1)}$ for this dressing. These minima are plotted versus the resepective mass ratio in Fig.~\ref{fig:mass-renorm} (c).
			}
		\label{fig:mass-renorm-p_out}
	\end{figure}

	Inserting this dispersion relation into momentum and energy conservation relations, see Eq.~\eqref{eq:A1_final_momentum}, we find a decrease of the critical momentum $\pc^{(1)}$ with increasing dressing, see Fig.~\ref{fig:mass-renorm-p_out}. The minima here are below $\pc^{(1)}$ and therefore redefine the critical momentum. These minima are plotted versus the resepective mass ratio in Fig.~\ref{fig:mass-renorm} (c).
	
\bibliography{bibliography}

\end{document}